\documentclass[twoside,twocolumn]{article}

\usepackage[hmarginratio=1:1,margin={1.45cm,1.45cm}, top=32mm,columnsep=20pt]{geometry} 

\usepackage{amsmath,graphicx,balance,amssymb}
\usepackage{balance}

\def\Y{\mathbf{Y}}

\usepackage{microtype} 
\usepackage[english]{babel} 

\usepackage[hang, small,labelfont=bf,up,textfont=it,up]{caption} 
\usepackage{booktabs} 
\usepackage{enumitem} 
\setlist[itemize]{noitemsep} 

\usepackage{abstract} 

\usepackage{titlesec} 
\renewcommand\thesection{\Roman{section}} 
\renewcommand\thesubsection{\roman{subsection}} 
\titleformat{\section}[block]{\large\scshape\centering}{\thesection.}{1em}{} 
\titleformat{\subsection}[block]{\large}{\thesubsection.}{1em}{} 

\usepackage{fancyhdr} 
\usepackage{titling} 

\usepackage{hyperref} 

\pretitle{\begin{center}\bfseries} 
\posttitle{\end{center}} 
\title{A RECURRENT ENCODER-DECODER APPROACH WITH SKIP-FILTERING CONNECTIONS FOR MONAURAL SINGING VOICE SEPARATION} 
\author{%
{Stylianos Ioannis Mimilakis$^{*}$, Konstantinos Drossos$^{\dagger}$, Tuomas Virtanen$^{\dagger}$, Gerald Schuller$^{*\ddag}$} \\[1ex] 
\normalsize $^{*}$Fraunhofer IDMT, Ilmenau, Germany \\ \{mis, shl\}@{idmt.fraunhofer.de} \\
\normalsize{$^{\dagger}$Tampere University of Technology, Tampere, Finland} \\ 
\normalsize{firstname.lastname@}{tut.fi} \\
\normalsize{$^{\ddag}$Technical University of Ilmenau, Ilmenau, Germany}
}
\date{} 
\renewcommand{\maketitlehookd}{%
\begin{abstract}
\noindent The objective of deep learning methods based on encoder-decoder architectures for music source separation is to approximate either ideal time-frequency masks or spectral representations of the target music source(s). The spectral representations are then used to derive time-frequency masks. In this work we introduce a method to directly learn time-frequency masks from an observed mixture magnitude spectrum. We employ recurrent neural networks and train them using prior knowledge only for the magnitude spectrum of the target source. To assess the performance of the proposed method, we focus on the task of singing voice separation. The results from an objective evaluation show that our proposed method provides comparable results to deep learning based methods which operate over complicated signal representations. Compared to previous methods that approximate time-frequency masks, our method has increased performance of signal to distortion ratio by an average of 3.8 dB.
\end{abstract}
\hspace{3.66cm} \begin{keywords}
	Music source separation, deep learning, denoising autoencoders
\end{keywords}
}

\begin{document}

\maketitle

\section{Introduction}
\label{sec:intro}
The under-determined separation of audio and music signals from mixtures is an active research area in the field of audio signal processing. The main objective is to estimate individual sources contained in an observed single-channel (i.e. monaural) mixture. An important task which has attracted a lot of attention is the estimation of singing voice and background music~\cite{sisec17}. The most widely-used strategy to achieve the estimation of individual sources employs time-varying filters, usually in the short-time Fourier transform (STFT) domain. These filters, henceforth denoted as time-frequency masks, are derived from rational models which incorporate prior information about the spectral representation of each source in the mixture~\cite{liutkus_alpha, ps_masks}.

When sources are already known (i.e. informed source separation)~\cite{iss_on_graph}, the process of source separation is almost trivial by employing an optimal time-frequency masking strategy~\cite{liutkus_alpha}. On the other hand, when the sources are not known, a prior estimation of each individual source takes place. To that aim, numerous approaches have been proposed. More specifically, in~\cite{cano} it is proposed to exploit phase information for separating sources that have harmonic and impulsive characteristics, while in~\cite{repet} the repetitive structure of music sources is studied for separating singing voice and background music. Alternative methods approximate the mixture magnitude via non-negative low rank modeling~\cite{cauchy_nmf, fitz_projet}, and/or matrix completion via robust principal component analysis (RPCA)~\cite{rpca}. 

Studies in music source separation have shown that supervised methods based on deep learning can yield state-of-the-art results in singing voice separation \cite{sisec17}. Such methods can be roughly discriminated into two main categories. In the first category the goal is to train deep neural networks, to approximate an ideal time-frequency mask from observed magnitude spectral representations~\cite{grais16, grais162, cha17}. The second category exploits denoising auto-encoders (i.e. neural networks trained to map from a noisy observation to a clean one), with the main goal of estimating the magnitude spectral representation of individual music sources from input mixtures~\cite{uhl15, huang, mim16}. Then, these estimates are combined to derive~\cite{uhl15, mim16, uhl17} or optimize~\cite{huang, nug16} a time-frequency mask that filters the input mixture. However, the quality of the time-frequency mask is heavily depended on the computation of the sources~\cite{huang, uhl17} and the time-frequency mask approximation is not part of the optimization process (only the estimation of the sources is). 

An exception to the latter is the work in~\cite{huang}. In that work, in order to predict the time-frequency mask the sources are first predicted and then combined in a generalized Wiener filtering process. Although, and according to~\cite{huang}, the approximation of the time-frequency mask is subject to optimization, this optimization is based on the ability of the previous neural network layers to estimate the sources. 

In this work we propose a method for predicting a time-frequency mask from the observed mixture magnitude spectrogram, and optimizing it according to its performance on estimating the target source. During training, the only prior knowledge is the mixture and the target source magnitude spectrograms. After training, only the mixture magnitude spectrogram is required. This approach differs from the existing ones in music source separation because: a) we do not base the prediction of the time-frequency mask on the prior estimation of the source(s), and b) we do not require a prior knowledge of the ideal time-frequency mask for training. We let a recurrent neural network (RNN) to predict the time-frequency mask. Then, based on that time-frequency mask, we estimate the magnitude spectrogram of the target source using skip-filtering connections, a highway network, and a generalized Wiener filtering process. The RNN and the highway network are jointly trained. 

The rest of the paper is organized as follows: Section~\ref{sec:proposedmethod} presents the proposed method, followed by Section~\ref{sec:III} which provides information about the training and evaluation of the proposed method. Section~\ref{sec:IV} presents the obtained results from the experimental procedure, followed by discussion. Conclusions are in Section~\ref{sec:V}.

\section{Proposed method}
\label{sec:proposedmethod}
\begin{figure}[!t]
	\centering
	\includegraphics[width=1.\columnwidth, keepaspectratio]{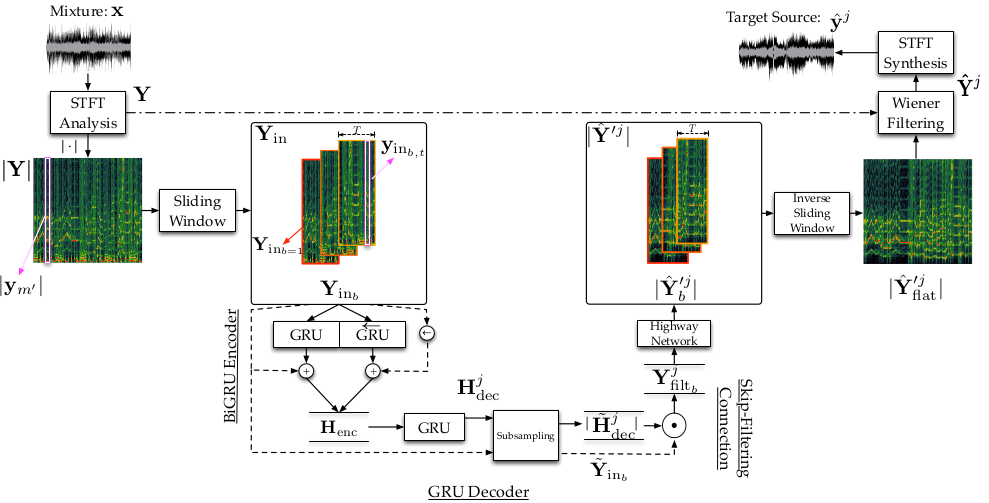}
	\caption{Illustration of the proposed encoder-decoder approach with skip-filter connections.}
	\label{fig:sys}
\end{figure}

Our method accepts as an input the time-domain samples of a monaural mixture vector $\mathbf{x}$ and produces the time-domain samples of the $j$-th target source vector $\mathbf{\hat{y}}^j$, as illustrated in Fig.~\ref{fig:sys}.
We calculate the matrix $\Y$, which contains the complex-valued time-frequency representation of $\mathbf{x}$. In order to use short sequences of frames with context information (i.e. previous and next frames), we create the tensor $\Y_{\text{in}}$ consisting of overlapping segments of the magnitude of $\Y$. We use each matrix in $\Y_{\text{in}}$ as an input to a single-layered, bi-directional gated recurrent unit (BiGRU), the encoder, in order to let our method learn temporal inter-dependencies of the input data. The output of the encoder is used as an input to a single layered GRU, the decoder, in order to estimate a time-frequency mask. We combine the estimated time-frequency mask and the input to the encoder using skip-filtering connections, in order to estimate the magnitude spectrogram of the target source, $\mathbf{Y}^{j}_{\text{filt}}$. The encoder and the decoder are optimized by minimizing the generalized Kullback-Leibler divergence, $\mathcal{L}_{KL}$, between the true and the estimated magnitude spectrogram of the target source.

The $\mathbf{Y}^{j}_{\text{filt}}$ is used as an input to a highway network in order to reduce the interferences from the rest sources. The encoder, the decoder, and the highway network are optimized using $\mathcal{L}_{KL}$ between the true magnitude spectrogram of the target source and the output of the highway network, plus an $L_2$ regularization term. The output of the highway network with the complex-valued mixture representation, $\Y$, are given as an input to a generalized Wiener filtering process. The latter produces the complex-valued time-frequency representation of the target source and further reduces the interferences from the rest sources. The final output, $\mathbf{\hat{y}}^j$, is calculated through an overlap and add synthesis procedure~\cite{gla}.

\subsection{Input processing}
Vector $\mathbf{x}$ is sliced in $M$ overlapping time frames of length $N$, using a shift/hop-size of $H$ samples. The overlapping time frames are element-wise multiplied by the Hamming windowing function. This yields a matrix of overlapping segments of the original mixture, from which the complex-valued time-frequency representation $\Y \in \mathbb{C}^{M \times N} $ is acquired by applying the short-time Fourier (STFT) analysis for all $M$ time segments. From the STFT representation of the mixture $\Y$, we compute the magnitude spectrum $|\Y|$, with $|\cdot|$ denoting the element-wise magnitude of the matrix. Using a sliding window over the time-frames of $|\Y|$ we form the tensor $\Y_{\text{in}} \in \mathbb{R}_{\geq 0}^{B \times T \times N}$, where $B=\lceil M / T \rceil$ and $T$ is an integer indicating the amount of STFT time-frames used for encoding. $\Y_{\text{in}}$ consists of a $B$ number of matrices  that encompass $T \times N$ segments of the mixture magnitude spectrogram $|\Y|$. Each matrix $\Y_{\text{in}_b} \in \mathbb{R}_{\geq 0}^{T \times N}$ is a sequence of $T$ overlapping time-frames, and is acquired from $|\Y|$ as
\begin{align}\label{eq:reshapeA}
\Y_{\text{in}_b} &= [|\mathbf{y}_{m' + 1}|, \ldots, |\mathbf{y}_{m' + T}|]\text{, where}\\
m' &= (b-1)\times T - (b-1)\times 2\times L\text{, }\label{eq:reshapeB}
\end{align}
\noindent
and $L$ and $m'$ are integers indicating the amount of time-frames that will be used as context information, and the time-frame location in $|\Y|$ ($|\mathbf{y}_{m'}|$), respectively.
In the case that $m' > M$, $|\mathbf{y}_{m'}|$ is a vector of zeros and of length $N$. The usage of overlapping time-frames is interpreted as having $2\times L$ context frames ($L$ before and $L$ after) for $T-(2\times L)$ length sequence of time-frames and is necessary for the sub-sampling operation, imposed by the decoding part of our proposed method. The term $2\times L$ is derived experimentally.

\subsection{Mask and source estimation}
Each $\Y_{\text{in}_b}$ is used as an input to the BiGRU encoder, which processes it according to the equations describing a GRU and using the proposed non-linearities~\cite{bahdanau:2015:iclr}. The BiGRU encoder consists of two GRUs, the forward GRU which accepts $\Y_{\text{in}_b}$ as an input, and the backward one which accepts ${\overleftarrow{\Y}_{\text{in}_b}} = [\mathbf{y}_{\text{in}_{b, T}}, \ldots, \mathbf{y}_{\text{in}_{b,t}}, \ldots, \mathbf{y}_{\text{in}_{b,1}}]\text{,}$ where $\mathbf{y}_{\text{in}_{b,t}} \in \mathbb{R}_{\geq 0}^{N}$ is a vector in $\Y_{\text{in}_b}$ at time-step $t$, and $\overleftarrow{ }$ denotes the direction of recursion over the time sequences. The hidden outputs from the bi-directional encoder at time-step $t$, $\mathbf{h}_{t}$ and $\overleftarrow{\mathbf{h}}_t$, are concatenated and updated using residual connections~\cite{skip}, as
\begin{equation}
\label{eq:concat}
{\mathbf{h}_{\text{enc}_t}} = [({\mathbf{h}}_{t} + {{\mathbf{y}}_{\text{in}_{b,t}}})^{\text T}, ({\overleftarrow{\mathbf{h}}_{t}} + {\overleftarrow{\mathbf{y}}_{\text{in}_{b,t}}})^{\text T} ]^{\text T}
\end{equation}
\noindent
and the output of the encoder is
\begin{equation}
\mathbf{H}_{\text{enc}} =[{\mathbf{h}_{\text{enc}_1}}, \ldots, 
{\mathbf{h}_{\text{enc}_t}}, \ldots,{\mathbf{h}_{\text{enc}_T}}]\text{.}
\end{equation}
$\mathbf{H}_{\text{enc}}$ is used as an input to the decoder, producing an output denoted as $\mathbf{H}_{\text{dec}}^j$. The subscripts {``enc''} and {``dec''} stand for encoder and decoder, respectively. The superscript $j$ signifies that the values of the matrix are dependent on the $j$-th source.  The concatenation imposed by Eq.~(\ref{eq:concat}) increases the dimensionality of the encoded information by a factor of two ($\mathbf{H}_{\text{enc}}$ is of shape ${T \times (N \times 2)}$). For this reason the decoder utilizes dimensionality reduction, such the $\mathbf{H}_{\text{dec}}^j$ matches the dimensionality of $\Y_{\text{in}_b}$. 
Furthermore, a sub-sampling operation over the time-steps of $\mathbf{H}_{\text{dec}}^j$ is applied, defined as
\begin{equation}
\label{eq:subsam}
\tilde{\mathbf{H}}_{\text{dec}}^j = [ {\mathbf{h}}^{j}_{\text{dec}_{1+L}}, {\mathbf{h}}^{j}_{\text{dec}_{2+L}}, 
\ldots, {\mathbf{h}}^{j}_{\text{dec}_{T-L}}],
\vspace{-1mm}
\end{equation}
where $L$ in Eq.~(\ref{eq:subsam}) indicates the amount of time steps to be discarded from the start and end of ${\mathbf{H}}_{\text{dec}}^j$. The intuition behind Eq.~(\ref{eq:subsam}) is to allow the flow of information from previous and preceding context frames $L$, without dramatically increasing the learning complexity of longer time interdependencies~\cite{subrnn}. After the sub-sampling operation, we apply the skip-filtering connections between $\mathbf{\tilde{H}}_{\text{dec}}^j$ and a sub-sampled version of the input data $\tilde{\mathbf{\Y}}_{\text{in}_{b}}$ as
\begin{align}
\label{eq:sfilt}
\Y_{\text{filt}_{b}}^j =& \tilde{\Y}_{\text{in}_{b}} \odot | \tilde{\mathbf{H}}_{\text{dec}}^j| \text{, where}\\
\tilde{\Y}_{\text{in}_{b}} =& [ {\mathbf{y}}_{\text{in}_{b, L}}, { }_{\cdots}, {\mathbf{y}}_{\text{in}_{b, T-L}}]\text{,}
\end{align}
\noindent
$\odot$ is the Hadamard product, $\tilde{\mathbf{H}}_{\text{dec}}^j$ is interpreted as a time-frequency mask specific to the source $j$, and $\tilde{\Y}_{\text{in}_{b}}$ is the sub-sampled version of $\Y_{\text{in}_{b}}$. According to Eq.~(\ref{eq:sfilt}), $\Y_{\text{filt}_{b}}^j$ can be considered as a filtered version of the input sequence $\tilde{\Y}_{\text{in}_{b}}$ approximating the magnitude spectrogram of the $j$-th target source, and thus the subscript ``filt''.  We minimize the $\mathcal{L}_{KL}$ between the true and estimated magnitude spectrogram of the $j$-th target source to train the encoder and the decoder.

The skip-filtering connections directly filter the input mixture magnitude sequence by the output of the decoder. The motivation behind this is to enforce $\tilde{\mathbf{H}}_{\text{dec}}^j$ to be a time-frequency mask derived through optimization, based on the prior knowledge of the magnitude spectrogram of the target source and not on the prior knowledge of an ideal time-frequency mask.
This is inspired by the denoising source separation framework presented in~\cite{den_ss}, where the separation of a stationary source in the frequency domain is achieved by learning a matrix which contains on its main diagonal the scalar values for each frequency sub-band (i.e. a filter). The same operation can be realized by the Hadamard product of vectors in the frequency domain, which can be implemented by the skip-filtering connections presented in this work and the speech separation approach presented in~\cite{wen14} which is denoted as signal approximation 
\footnote{At the time of publication the authors of this work were not aware of the speech separation method presented in~\cite{wen14}. To acknowledge the findings of~\cite{wen14} we have addressed the necessary changes in this manuscript version.}.

\subsection{Enhancement of estimated magnitude spectrogram}
In order to enhance the outcome of the filtering process of Eq.~(\ref{eq:sfilt})~\cite{mim16}, $\Y_{\text{filt}_{b}}^j$ is used as an input to a single layer of highway neural networks~\cite{hw15}, with shared weights in time, as
\begin{equation}
\begin{split}
\label{eq:highway}
|\hat{\Y}'^j_{b}| = \sigma(\Y^j_{\text{filt}_{b}} \, \mathbf{W}^{h} + 
\mathbf{b}^{h}) \odot g(\Y^j_{\text{filt}_{b}} \, \mathbf{W}^{tr} + \mathbf{b}^{tr}) + \\
\Y^j_{\text{filt}_{b}} \odot (1 - \sigma(\Y^j_{\text{filt}_{b}} \, \mathbf{W}^{tr} + 
\mathbf{b}^{tr}))\text{,}
\end{split}
\end{equation}
where $|\hat{\Y}'^j_{b}|$ is the enhanced estimated magnitude spectrogram of the $j$-th target source \cite{mim16}, $\sigma$ is the sigmoid function, $g$ is defined as $g(x) = \max(0, x)$, and $\mathbf{W}^{tr}$ and $\mathbf{W}^{h}$ are the weight matrices of the transformation and gating operations with their corresponding bias vectors $\mathbf{b}^{h}$ and $\mathbf{b}^{tr}$, respectively. All the layers are trained by minimizing the $\mathcal{L}_{KL}$ between the true and the enhanced estimated magnitude spectrogram of the $j$-th target source, plus an additional $L_2$ penalty, for regularizing $|\hat{\Y}'^j_{b}|$, scaled by $\lambda = 1e^{-4}$. Given that $L_2$ penalty is higher for non pseudo-periodic information~\cite{rpca}, we use it in order to enforce the highway network to enhance the estimated magnitude spectrogram of the singing voice. 

\subsection{Final output and implementation details}
We iterate through all the above presented equations for all $b$ in $\Y_{\text{in}_{b}}$ and the result is the tensor $|\hat{\Y}'^j|\in \mathbb{R}_{\geq 0}^{B\times T'\times N}$, where $T'=T - 2\times L$. Then, $|\hat{\Y}'^j|$ is reshaped to $|\hat{\Y}'^j_{\text{flat}}|\in \mathbb{R}_{\geq 0}^{M\times N}$, by reversing the process in Eq.~(\ref{eq:reshapeA}) and~(\ref{eq:reshapeB}). The subscript {``flat''} is used to refer to the aforementioned reshaping procedure. The final output of the proposed method is the complex-valued spectral representation of the $j$-th source $\hat{\Y}^{j} \in \mathbb{C^{M\times N}}$ computed as
\begin{align}
\label{eq:final_1}
\hat{\Y}^{j} &= \mathbf{M}^{j} \odot \Y\text{, where} \\
\label{eq:final_2}
{\mathbf{M}}^{j} &= \frac{| {\hat{\Y}'^j_{\text{flat}}} | ^{\alpha}}{|{\Y}|^{\alpha}}\text{,}
\end{align}
$\alpha \in (0, 2]$ is an exponent applied element-wise, signifying the statistical assumption(s) about the sources and their additive property in the magnitude time-frequency domain~\cite{liutkus_alpha}. Furthermore, the division is also performed element-wise. The reasoning behind Eqs.~(\ref{eq:final_1})--(\ref{eq:final_2}) and the $\alpha$ factor, is to infer the information that the mixture encapsulates about the phase and the magnitude of the target source, practically improving the interference reduction from other concurrently active sources. The reconstruction of the time-domain function of the target source $\mathbf{\hat{y}}^j$ is achieved by the STFT synthesis operation, for all M time segments, followed by the overlap and add method presented in~\cite{gla}.

The highway network and the GRUs in the encoder consist of 1025 neurons each and the GRU in the decoder has 2050 neurons. This results in approximately 24 million parameters in total for our method. All weight matrices 
are randomly initialized using the method presented in~\cite{glorot} and jointly trained using the adam algorithm~\cite{adam}, with an initial learning rate of $1e^{-3}$, over $b$ batches of 16, and an $L_2$ based gradient norm clipping equal to 0.35. The training is terminated if after two consecutive iterations over all the available training batches, no minimization took place. The implementation of the proposed method is based on the keras~\cite{keras} and Theano~\cite{theano} frameworks. The above parameters are chosen experimentally by informal listening tests and subjectively evaluating the obtained quality of separation, with data drawn from the development subset of DSD$100$ (presented in the next section).

\section{Evaluation}
\label{sec:III}
\subsection{Dataset and preprocessing}
In order to assess the performance of the proposed method we focus on the task of singing voice and background music separation. The Demixing Secret Dataset (DSD$100$)\footnote{\url{http://www.sisec17.audiolabs-erlangen.de}} is used for training and evaluating our approach. DSD$100$ consists of $100$ professionally produced multi-tracks of various music genres, sampled at $44.1$kHz. The dataset is by default divided evenly into development and evaluation subsets, each consisting of $50$ multi-tracks, and each multi-track contains four stereo target sources forming the produced mixture.

For each multi-track contained in the development subset, a monaural version of each of the four sources is generated by averaging the two available channels. Afterwards, two signals are generated. One containing the corresponding monaural mixture of all the monaural sources (singing voice, bass, drums, etc) and a second containing the mixture of all the monaural sources but the singing voice. The mixing gain values for each source are not modified (no data augmentation is applied). An analysis operation using the STFT is applied to the two preceding mixture signals and the target singing voice, using $H=256$ and $N' = 2048$. Since we are concerned with real-valued signals, their time-frequency representation using the DFT is Hermitian and thus the redundant information is discarded, resulting into a dimensionality of $N = 1025$. The length of the sequences is set to $T = 18$, modeling approximately $120$ ms, and $L = 3$. To avoid inconsistencies due to the sub-sampling operation of Eq.~(\ref{eq:subsam}), the considered shifts for acquiring $\Y_{\text{in}}$ are overlapping by six time-frames ($2 \times L$). During training, the true source $|\mathbf{Y}^j|$ is the outcome of a generalized Wiener filtering using the a priori knowledge from the dataset for each source and a value of $\alpha = 1$.

\subsection{Metrics and evaluation procedure}
To evaluate our method we use the standard metrics employed in the music source separation evaluation campaign~\cite{sisec17}. These are the signal to distortion ratio (SDR) and the signal to interference ratio (SIR). These metrics are computed by using the output of the proposed method on the evaluation sub-set and the true sources acquired from the multi-tracks of the same sub-set. Results from three different strategies using the proposed method are reported. For the first one, we train our method using the magnitude spectrogram of the singing voice $|\mathbf{Y}^{j=1}|$. After the training, we obtain an estimation of the singing voice using the input mixture spectrogram and an $\alpha$ value equal to $1.7$, as according to Eq.~(\ref{eq:final_1}) and~(\ref{eq:final_2}). For the rest two strategies, we use two deep neural networks trained separately. Both of them are based on our method. One to predict the singing voice $|\mathbf{Y}^{j=1}|$ and another to predict the background music without the singing voice $|\mathbf{Y}^{j=2}|$. The predicted magnitude spectral estimates (i.e. $|\mathbf{\hat{Y}}^{j=1}|$ and $|\mathbf{\hat{Y}}^{j=2}|$) are used in the Eq.~(\ref{eq:final_2}) and the Eq.~(\ref{eq:final_1}) to compute the final estimation of the $|\mathbf{Y}^{j=1}|$. In Eq.~(\ref{eq:final_2}), the numerator is defined as $|\mathbf{\hat{Y}}^{j=1}|^\alpha$ and we replace the $|\mathbf{Y}|^\alpha$ of the denominator with $|\hat{\Y}|^\alpha = |\mathbf{\hat{Y}}^{j=2}|^\alpha + |\mathbf{\hat{Y}}^{j=1}|^\alpha$. For the second strategy we use $\alpha=1.7$ and for the third $\alpha=2$. We will refer to these three strategies as GRU-S (the first), GRU-D (the second), and GRU-DWF (the third). The values for $\alpha$ are chosen according to the generalized Wiener filtering (i.e. $\alpha = 2$) and its extension to heavy-tailed distributions~\cite{liutkus_alpha}.

We compare with state-of-the-art methods dealing with monaural singing voice separation. More specifically, we compare the results from our proposed method against: i) two approaches based on deep feed-forward neural networks of approximately $17$ million parameters each, trained using data augmentation to predict both binary and soft time-frequency masks derived from ~\cite{ps_masks}, denoted as GRA2 and GRA3~\cite{grais16} ii) a convolutional based encoder-decoder approach with approximately $27$ million parameters, trained using data augmentation to predict a soft time-frequency mask, denoted as CHA~\cite{cha17}, and iii) two additional deep denoising auto-encoders operating on the common fate signal representation~\cite{cmf} denoted as STO1, STO2. As for the oracle estimation, the results from ideal binary masking (IBM) are also presented. We report results for the overall performance on the employed dataset and for three clusters of music genres, namely Jazz/Pop/Rock, Electronic/Rap, and Heavy Metal. 

\section{Results \& Discussion}
\label{sec:IV}
\begin{figure}[!t]
	\centering
	\includegraphics[width=1.\columnwidth, keepaspectratio]{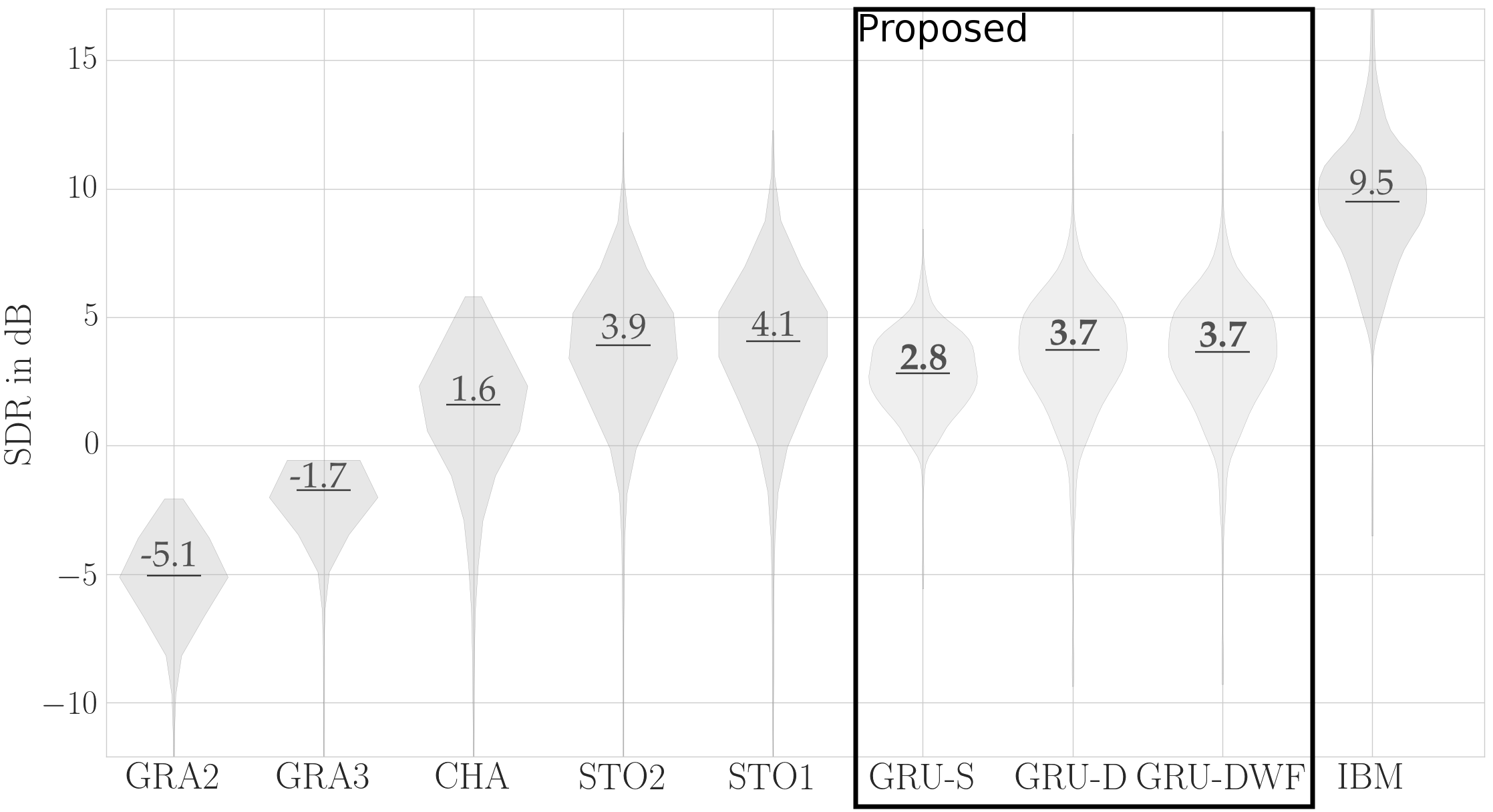}
	\caption{Analysis of variance of SDR for previous approaches and the proposed ones. Lines denote the median values.}
	\label{fig:sdr}
\end{figure}

\begin{figure}[!t]
	\centering
	\includegraphics[width=1.\columnwidth, keepaspectratio]{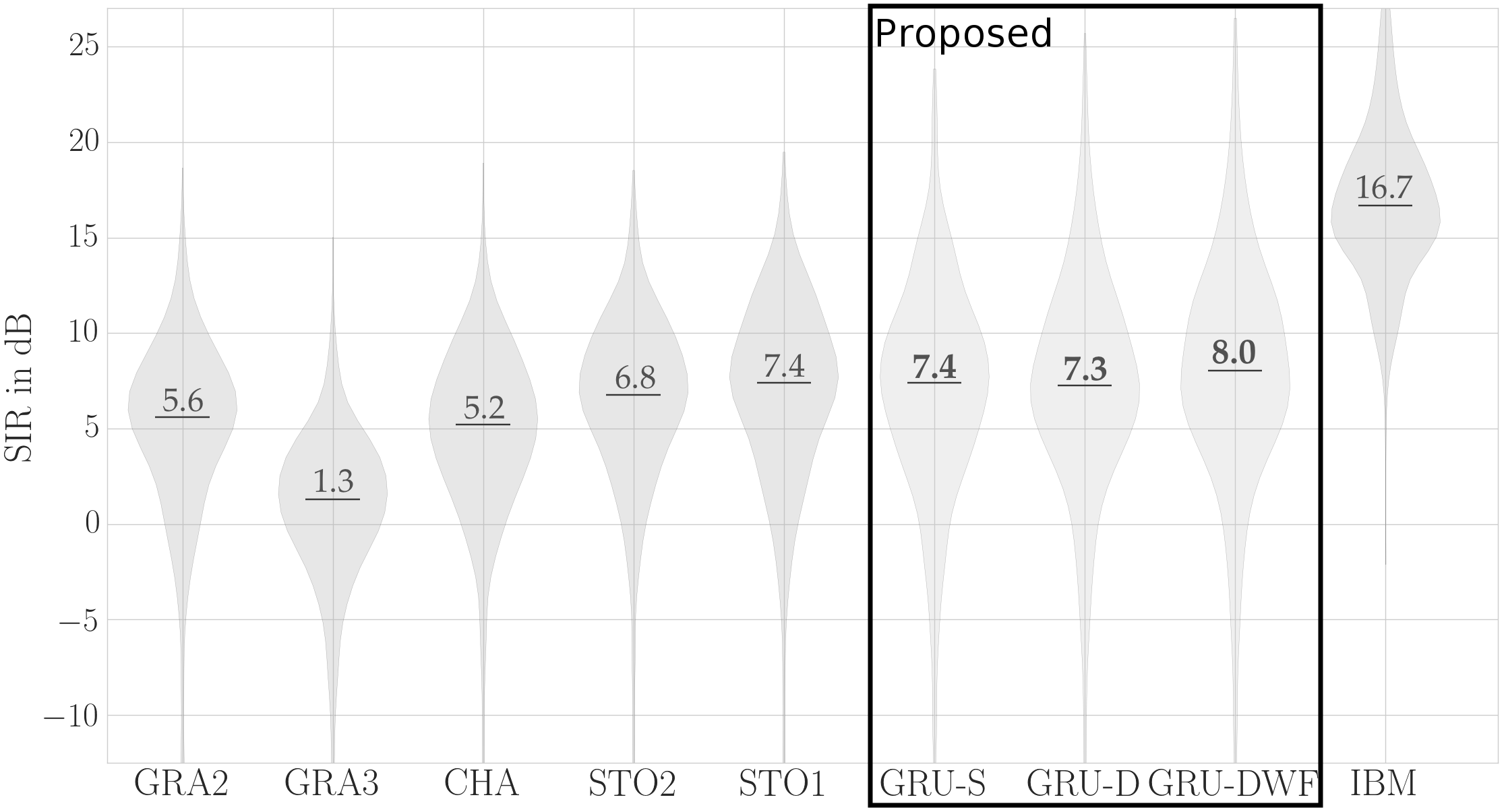}
	\caption{Analysis of variance of SIR for previous approaches and the proposed ones. Lines denote the median values.}
	\label{fig:sir}
\end{figure}

\begin{figure}[!t]
	\centering
	\includegraphics[width=1.\columnwidth, keepaspectratio]{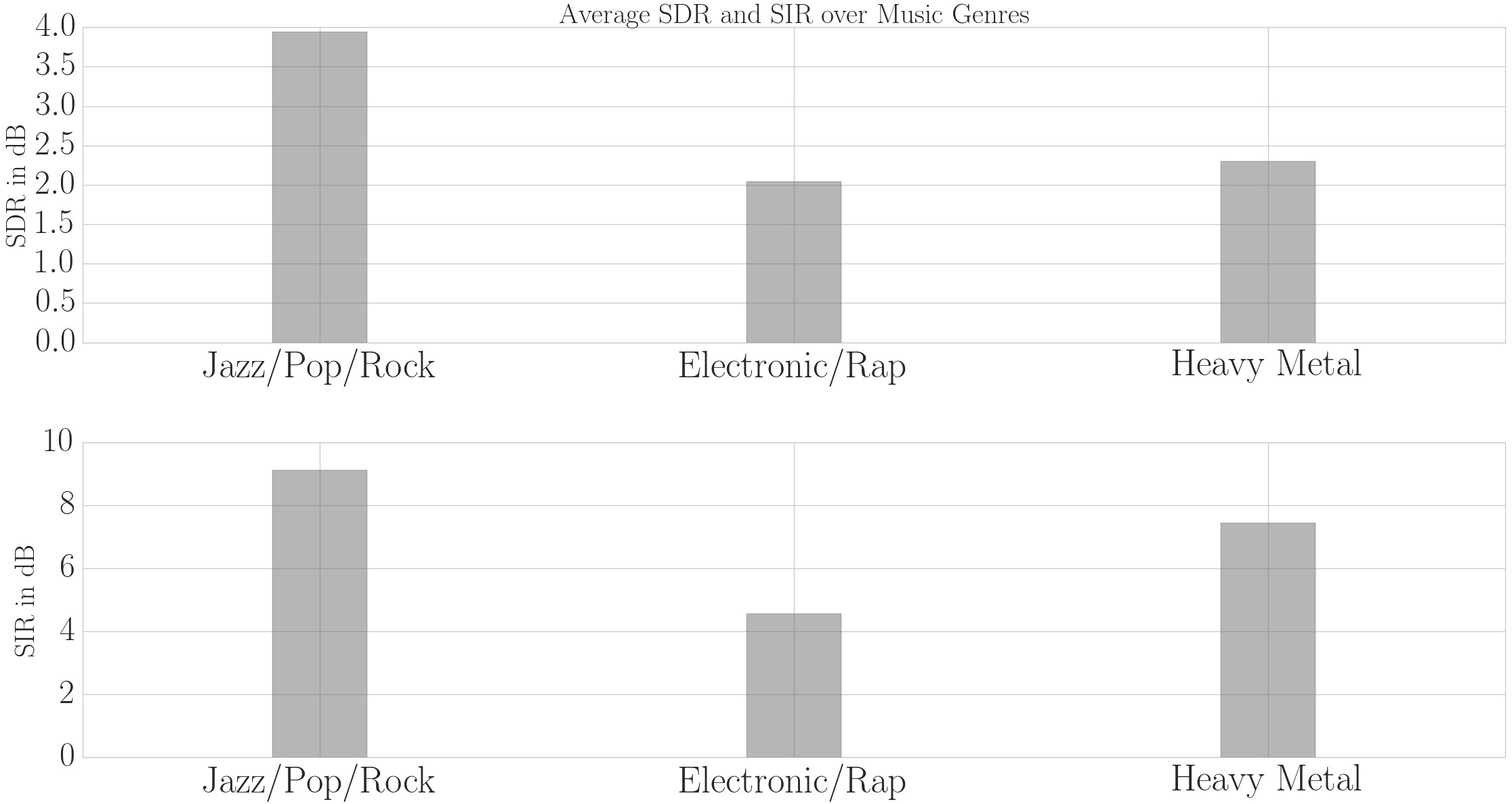}
	\caption{Average objective performance of the three presented strategies over three clusters of music genres.}
	\label{fig:genre}
\end{figure}

Results from the SDR and SIR metrics for the above methods and the oracle estimation are illustrated in Figures~\ref{fig:sdr} and~\ref{fig:sir}. Examining the differences in SDR and SIR metrics between the methods are trained to approximate the ideal time-frequency masks (GRA3, CHA) and GRU-D, it can be observed that the median SDR has been improved by approximately $5.4$ dB when compared to GRA3 and by $2.1$ dB when compared to CHA. An average improvement of $3.5$ dB for the SIR can be observed regarding the methods that predict of the time-frequency mask (GRA2, GRA3, CHA). On the other hand, when it comes to comparison with common denoising auto-encoders (STO1, STO2) and our method, a marginal average loss of $0.3$ dB in the median SDR and a gain of $0.25$ dB and $0.9$ dB in the median SIR can be observed according to the results over the GRU-D and GRU-DWF cases, respectively. We believe (but we don't have any evidence) that these marginal differences can be attributed to the more sophisticated signal representations, such as the common fate model~\cite{cmf}, used in STO1, STO2. In contrast, our method operates on top of a magnitude spectral representation computed from a STFT. 
By inspecting the differences between the proposed strategies GRU-S and GRU-D, it can be seen that SDR can be increased by approximately $0.9$ dB by incorporating separate deep neural networks for approximating additional sources contained in mixtures. On the other hand, dramatic differences in SIR between the three strategies were not observed, unless the value for $\alpha$ was increased, like in the case of GRU-DWF. This shows, that combining multiple deep neural networks in generalized Wiener filtering leads to improved SDR. Additionally, the additivity property, acknowledged in generalized Wiener filtering~\cite{liutkus_alpha}, of the estimates of deep neural networks might not hold true without an explicit cost objective. Nonetheless, it can used for improving the interference reduction. In Figure~\ref{fig:genre} are the average SDR and SIR measures of the three employed strategies over the three employed music genres.

As it can be seen, the performance over the cluster Jazz/Pop/Rock is higher than the performance of for the other clusters. An explanation to this is that the DSD$100$ dataset contains training multi-tracks mainly from pop and rock music genres. This means that the poorer performance of our method can be attributed to the fact that recurrent models might need additional data for modeling more complicated structures of singing voice.

\section{Conclusions \& Future Work}
\label{sec:V}
In this work we presented a deep learning method for music source separation. The method used an encoder-decoder configuration based on GRUs and skip-filtering connections between input spectral representations and their hidden latent variables, forcing the GRUs to approximate a time-frequency masking operation. Its application to monaural singing voice separation was studied and assessed objectively. The obtained results signify that the skip-filtering connections can be used for approximating time-frequency masks, providing comparable results to state of the art deep learning approaches. Future work will focus on psycho-acoustically motivated loss minimization and exploring sparsity priors for improving the approximated time-frequency mask. Subjective assessment of the plausible extensions of our methodology are also emerging~\cite{cano2}. Source code and listening examples can be found under: {\url{https://github.com/Js-Mim/mlsp2017_svsep_skipfilt}}.
\section{Acknowledgements}
We would like to thank the authors of~\cite{sisec17} for making the results of the evaluation available. The major part of the research leading to these results has received funding from the European Union's H2020 Framework Programme (H2020-MSCA-ITN-2014) under grant agreement no 642685 MacSeNet. Konstantinos Drossos was partially funded from the European Union's H2020 Framework Programme through ERC Grant Agreement 637422 EVERYSOUND. Minor part of the computations leading to these results were performed on a TITAN-X GPU donated by NVIDIA.





\begin{thebibliography}{99} 
\small
\balance
\bibitem{sisec17}
A.~Liutkus, F.-R. St{\"o}ter, Z.~Rafii, D.~Kitamura, B.~Rivet, N.~Ito, N.~Ono,
and J.~Fontecave,
\newblock ``The 2016 signal separation evaluation campaign,''
\newblock in {\em Latent Variable Analysis and Signal Separation: 13th
	International Conference, LVA/ICA 2017}, 2017, pp. 323--332.

\bibitem{liutkus_alpha}
A.~Liutkus and R.~Badeau,
\newblock ``Generalized wiener filtering with fractional power spectrograms,''
\newblock in {\em 40th International Conference on Acoustics, Speech and Signal
	Processing (ICASSP 2015)}, April 2015, pp. 266--270.

\bibitem{ps_masks}
H.~Erdogan, J.~R. Hershey, S.~Watanabe, and J.~Le Roux,
\newblock ``Phase-sensitive and recognition-boosted speech separation using
deep recurrent neural networks,''
\newblock in {\em 40th International Conference on Acoustics, Speech and Signal
	Processing (ICASSP 2015)}, April 2015, pp. 708--712.

\bibitem{iss_on_graph}
G.~Puy, A.~Ozerov, N.~Q.~K. Duong, and P.~Perez,
\newblock ``{Informed source separation via compressive graph sampling},''
\newblock in {\em {42nd International Conference on Acoustics, Speech and
		Signal Processing (ICASSP 2017)}}, Mar 2017, pp. 1--5.

\bibitem{cano}
E.~Cano, M.~Plumbley, and C.~Dittmar,
\newblock ``{Phase-based harmonic percussive separation},''
\newblock in {\em Proceedings of the Annual Conference of the International
	Speech Communication Association (Interspeech)}, Sept. 2014, pp. 1628--1632.

\bibitem{repet}
Z.~Rafii and B.~Pardo,
\newblock ``Repeating pattern extraction technique (repet): A simple method for
music/voice separation,''
\newblock {\em IEEE Transactions on Audio, Speech, and Language Processing},
vol. 21, no. 1, pp. 73--84, Jan 2013.

\bibitem{cauchy_nmf}
A.~Liutkus, D.~Fitzgerald, and R.~Badeau,
\newblock ``Cauchy nonnegative matrix factorization,''
\newblock in {\em Applications of Signal Processing to Audio and Acoustics
	(WASPAA 2015)}, Oct 2015, pp. 1--5.

\bibitem{fitz_projet}
D.~Fitzgerald, A.~Liutkus, and R.~Badeau,
\newblock ``{PROJET - Spatial Audio Separation Using Projections},''
\newblock in {\em {41st International Conference on Acoustics, Speech and
		Signal Processing (ICASSP 2016)}}, 2016, pp. 36--40.

\bibitem{rpca}
I.-Y. Jeong and K.~Lee,
\newblock ``Singing voice separation using {RPCA} with weighted $l_1$-norm,''
\newblock in {\em Latent Variable Analysis and Signal Separation: 13th
	International Conference, LVA/ICA 2017}, 2017, pp. 553--562.

\bibitem{grais16}
E.-M. Grais, G.~Roma, A.J.R. Simpson, and M.-D. Plumbley,
\newblock ``Single-channel audio source separation using deep neural network
ensembles,''
\newblock in {\em Audio Engineering Society Convention 140}, May 2016.

\bibitem{grais162}
E.-M. Grais, G.~Roma, A.J.R. Simpson, and M.-D. Plumbley,
\newblock ``Combining mask estimates for single channel audio source separation
using deep neural networks,''
\newblock in {\em Proceedings of the 17th Annual Conference of the
	International Speech Communication Association (Interspeech)}, Sept. 8-12
2016, pp. 3339--3343.

\bibitem{cha17}
P.~Chandna, M.~Miron, J.~Janer, and E.~G\'{o}mez,
\newblock ``Monoaural audio source separation using deep convolutional neural
networks,''
\newblock in {\em Latent Variable Analysis and Signal Separation: 13th
	International Conference, LVA/ICA 2017}, 2017, pp. 258--266.

\bibitem{uhl15}
S.~Uhlich, F.~Giron, and Y.~Mitsufuji,
\newblock ``Deep neural network based instrument extraction from music,''
\newblock in {\em 40th International Conference on Acoustics, Speech and Signal
	Processing (ICASSP 2015)}, 2015, pp. 2135--2139.

\bibitem{huang}
P.-S. Huang, M.~Kim, M.~Hasegawa-Johnson, and P.~Smaragdis,
\newblock ``Joint optimization of masks and deep recurrent neural networks for
monaural source separation,''
\newblock {\em IEEE/ACM Transactions on Audio, Speech, and Language
	Processing}, vol. 23, no. 12, pp. 2136--2147, Dec 2015.

\bibitem{mim16}
S.-I. Mimilakis, E.~Cano, J.~Abe{\ss}er, and G.~Schuller,
\newblock ``New sonorities for jazz recordings: Separation and mixing using
deep neural networks,''
\newblock in {\em Audio Engineering Society 2nd Workshop on Intelligent Music
	Production}, 2016.

\bibitem{uhl17}
S.~Uhlich, M.~Porcu, F.~Giron, M.~Enenkl, T.~Kemp, N.~Takahashi, and
Y.~Mitsufuji,
\newblock ``Improving music source separation based on deep neural networks
through data augmentation and network blending,''
\newblock in {\em 42nd International Conference on Acoustics, Speech and Signal
	Processing (ICASSP 2017)}, 2017, pp. 261--265.

\bibitem{nug16}
A.-A. Nugraha, A.~Liutkus, and E.~Vincent,
\newblock ``Multichannel music separation with deep neural networks,''
\newblock in {\em 24th European Signal Processing Conference (EUSIPCO)}, Aug
2016, pp. 1748--1752.

\bibitem{gla}
D.~Griffin and J.~Lim,
\newblock ``Signal estimation from modified short-time {F}ourier transform,''
\newblock {\em IEEE Transactions on Acoustics, Speech, and Signal Processing},
vol. 32, no. 2, pp. 236--243, Apr 1984.

\bibitem{bahdanau:2015:iclr}
D.~Bahdanau, K.~Cho, and Y.~Bengio,
\newblock ``Neural machine translation by jointly learning to align and
translate,''
\newblock in {\em International Conference on Learning Representations (ICLR)},
2015.

\bibitem{skip}
Y.~Wu et~al,
\newblock ``Google's neural machine translation system: Bridging the gap
between human and machine translation,''
\newblock {\em CoRR}, vol. abs/1609.08144, 2016.

\bibitem{subrnn}
A.~Graves,
\newblock {\em Hierarchical Subsampling Networks}, pp. 109--131,
\newblock Springer Berlin Heidelberg, Berlin, Heidelberg, 2012.

\bibitem{den_ss}
J.~S\"{a}rel\"{a} and H.~Valpola,
\newblock ``Denoising source separation,''
\newblock {\em J. Mach. Learn. Res.}, vol. 6, pp. 233--272, Dec. 2005.

\bibitem{wen14}
F.~Weninger, J.~R.~Hershey and J.~Le Roux and B.~Schuller
\newblock ``Discriminatively trained recurrent neural networks for single-channel speech separation,''
\newblock in {\em Proceedings of the IEEE Global Conference on Signal and Information Processing (GlobalSIP 2014)}, 2014, pp. 577--581.

\bibitem{hw15}
R.-K. Srivastava, K.~Greff, and J.~Schmidhuber,
\newblock ``Highway networks,''
\newblock {\em CoRR}, vol. abs/1505.00387, 2015.

\bibitem{glorot}
X.~Glorot and Y.~Bengio,
\newblock ``Understanding the difficulty of training deep feedforward neural
networks,''
\newblock in {\em In Proceedings of the International Conference on Artificial
	Intelligence and Statistics (AISTATS'10)}, 2010, pp. 249--256.

\bibitem{adam}
D.-P. Kingma and J.~Ba,
\newblock ``Adam: {A} method for stochastic optimization,''
\newblock {\em CoRR}, vol. abs/1412.6980, 2014.

\bibitem{keras}
F.~Chollet et~al.,
\newblock ``Keras (version 1.2.2),'' \url{https://github.com/fchollet/keras},
2015.

\bibitem{theano}
Theano~Development Team,
\newblock ``Theano: {A} python framework for fast computation of mathematical
expressions (version 0.9),''
\newblock {\em CoRR}, vol. abs/1605.02688, 2016.

\bibitem{cmf}
F.-R. St{\"o}ter, A.~Liutkus, R.~Badeau, B.~Edler, and P.~Magron,
\newblock ``Common fate model for unison source separation,''
\newblock in {\em International Conference on Acoustics, Speech and Signal
	Processing (ICASSP 2016)}, 2016, pp. 126--130.

\bibitem{cano2}
E.~Cano, D.~Fitzgerald, and K.~Brandenburg,
\newblock ``Evaluation of quality of sound source separation algorithms: Human
perception vs quantitative metrics,''
\newblock in {\em 24th European Signal Processing Conference (EUSIPCO)}, Aug
2016, pp. 1758--1762.
 
\end{thebibliography}
\end{document}